\documentclass[aps,prc,amsfonts,preprintnumbers,superscriptaddress,showpacs,nofootinbib,10pt]{revtex4-2}
\usepackage{epsfig}
\usepackage{graphics}
\usepackage{amsmath}
\usepackage{amssymb}
\usepackage{mathtools}
\usepackage{braket}
\usepackage{dsfont}
\usepackage{bm}
\usepackage[colorlinks=true,linkcolor=blue,urlcolor=blue,citecolor=blue]{hyperref}

\begin{document}
\title{Renormalizability and nonrenormalizability of nonlocal potentials}

\author{N.~Jacobi}
\email[]{Email: s83njaco@uni-bonn.de}
\affiliation{Universit\"at Bonn, D-53113 Bonn, Germany}

\author{A.~M.~Gasparyan}
\email[]{Email: ashot.gasparyan@rub.de}
\affiliation{Ruhr-Universit\"at Bochum, Fakult\"at f\"ur Physik und
        Astronomie, Institut f\"ur Theoretische Physik II,  D-44780
        Bochum, Germany}
      
\author{E.~Epelbaum}
\email[]{Email: evgeny.epelbaum@rub.de}
\affiliation{Ruhr-Universit\"at Bochum, Fakult\"at f\"ur Physik und
        Astronomie, Institut f\"ur Theoretische Physik II,  D-44780
        Bochum, Germany}

\begin{abstract}
We consider separable toy models of the nucleon-nucleon interaction
inspired  by chiral effective field theory. We show that nonlocality of the long-range forces causes the need
for nonlocal counter terms, or even makes the whole approach nonrenormalizable.
\end{abstract}

\maketitle
\section{Introduction}
Effective field theory (EFT) methods such as,
in particular, chiral EFT have become an indispensable tool for
analyzing few- and many-nucleon systems in a systematic fashion in
harmony with the symmetries of the Standard Model. 
The original idea to apply chiral EFT in the few-nucleon sector goes
back to the seminal works of Weinberg~\cite{Weinberg:1990rz,Weinberg:1991um} and has been further developed in numerous studies,
see Refs.~\cite{Bedaque:2002mn,Epelbaum:2008ga,Machleidt:2011zz,Epelbaum:2019kcf,Hammer:2019poc} for reviews.

Most of the practical applications require the introduction of  
a finite regulator, i.e., a momentum cutoff $\Lambda$ of the order of the 
EFT breakdown scale $\Lambda_b$, when solving the nuclear $A$-body
problem. 
An important ingredient of any field theory is renormalization, 
i.e.,~expressing observable quantities in terms of renormalized 
coupling constants $C^r_i$ instead of the bare ones appearing in the effective Lagrangian $C_i$.
The bare and renormalized coupling constants differ by counter terms, $C_i=C^r_i+\delta C_i$,
which absorb divergent and power-counting-breaking (PCB) contributions
caused by an interplay of various hard scales, such as a cutoff.
After renormalization, the calculated observables are expressed in
terms of renormalized constants and assumed to fulfill 
the employed power counting scheme.
In the case of chiral EFT, power counting corresponds
to an expansion in the small parameter $Q=q/\Lambda_b$, where
the soft scale $q$ is given by the pion mass $M_\pi$ and 
the external momenta. 
The issue of explicit renormalization of nuclear chiral EFT, being
very important for a justification of the corresponding practical applications, 
has not been fully resolved until recently.
First convincing results were obtained in Refs.~\cite{Gasparyan:2021edy,Gasparyan:2023rtj},
where renormalizability of nuclear chiral EFT in the two-nucleon
sector was explicitly demonstrated at leading (LO) and the next-to-leading orders (NLO).
Interestingly, renormalizability turns out \emph{not} to be a universal property
of quantum-mechanical interaction models, but rather a 
specific feature of the interactions derived from the effective (in particular) chiral Lagrangian.
In Ref.~\cite{Gasparyan:2025dpj}, various criteria of renormalizability are reviewed.
One of these criteria is related to the possibility of removing contributions
of the loop diagrams that involve positive powers of the cutoff $\Lambda$,
stemming from large-momentum (of the order of $\Lambda$) integration regions,
by an appropriate redefinition of the lower-order low-energy constants (LECs).
In chiral EFT, this requirement is fulfilled thanks to a specific structure of the effective two-nucleon potential.
The long-range part of the potential originating from the single- and
multi-pion exchange diagrams
is local. In other words, the singular part of the corresponding
potentials depends on the momentum
transfer $\vec q=\vec p\,'-\vec p$, where $\vec p$ ($\vec p\,'$) is the initial (final) center-of-mass
nucleon momentum, and not on $\vec p$ and $\vec p\,'$ individually.
This leads to certain bounds on the potential and its subtraction remainders as discussed in the next section.

In this note, we illustrate how the renormalization procedure works and 
how it could fail by considering interactions that do/do not possess the property of locality of the long-range forces.
We analyze several instructive examples based on simple separable $S$-wave potentials
motivated by nuclear chiral EFT.
Such interactions are often used as toy models in the literature to 
demonstrate some qualitative features of the realistic interactions
or as phenomenological models,
see, e.g., Refs.~\cite{Epelbaum:2009sd,Epelbaum:2017byx,Peng:2021pvo,Arellano:2024fym}.

Our paper is organized as follows.
In Sec.~\ref{Sec:Bounds}, the above-mentioned bounds for the chiral EFT potential
are reviewed. In Sec.~\ref{Sec:ToyModels} we present the toy models to be analyzed.
Their renormalization at NLO is discussed in Sec.~\ref{Sec:Renormalization}.
The main results of our paper are briefly summarized in Sec.~\ref{Sec:summary}.

\section{Bounds on the effective potential in chiral EFT}\label{Sec:Bounds}
We first review the bounds on the effective potential derived in ~\cite{Gasparyan:2021edy}
in order to prove renormalizability of nuclear chiral EFT in the two-nucleon sector.
The nucleon-nucleon ($NN$) potential in the plain-wave basis is assumed to be a finite sum of various structures $X_\alpha$:
\begin{align}
	V(\vec p\,',\vec p)=\sum_\alpha X_{\alpha}(\vec p\,',\vec p).\label{Eq:EFT_potential_structure}
\end{align}
Each structure $X_\alpha$ can be a product of any number
of the following substructures:
\begin{itemize}
	\item Static pion propagator corresponding to a single-pion exchange:
	\begin{align}
		f_{1\pi}(\vec q\,^2)=\frac{1}{\vec q\,^2+M_\pi^2}.
		\label{Eq:f_mu}
	\end{align}
	\item Multi-pion-exchange loop functions expressed via dispersive integrals:
	\begin{align}
		f_{n\pi}(\vec q\,^2)=\int_{n M_\pi}^\infty  \frac{\rho_n(\mu)d\mu}{\vec q\,^2+\mu^2},
	\end{align}
	where $n \geq 2$ and $\rho_n (\mu)$ denote the corresponding
        spectral function.   
	\item Homogeneous polynomial of momenta of degree $m$:
	\begin{align}
		& Q_m(p',p)=\sum_{\alpha,\beta}
		\mathcal{M}_{\alpha\beta}p^\alpha p'^\beta\,, \quad \alpha+\beta=m\,,
		\label{Eq:polynomial_Qm}
	\end{align}
	like, e.g., $\vec q\,^2$, $q_i q_j$ or $\vec p\times\vec p\,'$.
	\item Local and nonlocal regulators $F_{\Lambda}(p^2)$, $F_{\Lambda}(p'^2)$ and $F_{\Lambda}(\vec q\,^2)$,
	where $F_{\Lambda}(p^2)$ is either a power-like
	\begin{align}
		F_{\Lambda,m}(p^2)=\left(\frac{\Lambda^2}{p^2+\Lambda^2}\right)^m\,,
	\end{align}
	or Gaussian form factor
	\begin{align}
		F_{\Lambda,\text{Gauss}}(p^2) = \exp\left(-p^2/\Lambda^2\right).
	\end{align}
\end{itemize}
This form of the $NN$ potential covers essentially all types of 
the interactions derived from the chiral Lagrangian and used in the literature.

Assuming the above structure of the potential, one can derive the
following bounds:
\begin{align}
	&\left|V_i(\vec p\,',\vec p\,)\right|\le \frac{\mathcal{M}_{V_i}}{4\pi}
	V_{i,\text{max}}(p',p),\;\;\text{ if }  p'>p,\nonumber\\
	&\left|V_i(\vec p\,',\vec p\,)\right|
	\le \frac{\mathcal{M}_{V_i}}{4\pi}V_{i,\text{max}}(p,p'), \;\; \text{ if } p>p',
	\label{Eq:bound_full_LO}
\end{align}
for the EFT orders $i=0,2,\dots$.
The quantities $V_{i,\text{max}}(p',p)$ are some simple functions of
$p$ and $p'$, see Refs.~\cite{Gasparyan:2021edy,Gasparyan:2023rtj} for details. 

One can also deduce upper bounds for the subtraction remainders defined as:
\begin{align}
	\Delta_p^{(n)} f(p',p)&= 
	f(p',p)-\sum_{i=0}^{n}\frac{\partial^i f(p',p)}{i!\partial p^i}\bigg|_{p=0}p^i\,,\nonumber\\
	\Delta_{p'}^{(n)} f(p',p)&= 
	f(p',p)-\sum_{i=0}^{n}\frac{\partial^i f(p',p)}{i!\partial p'^i}\bigg|_{p'=0}(p')^i.
	\label{Eq:remainders}
\end{align}
The corresponding bounds are given by
\begin{align}
	&\left|\Delta_p^{(n)} V_i(\vec p\,',\vec p\,)\right| \le \frac{\mathcal{M}_{\Delta V_i,n}}{4\pi}
	\left(\frac{p}{p'}\right)^{n+1} V_{i,\text{max}}(p',p)\text{ if } p'>p,\nonumber\\
	&\left|\Delta_{p'}^{(n)} V_i(\vec p\,',\vec p\,)\right| \le \frac{\mathcal{M}_{\Delta V_i,n}}{4\pi}
	\left(\frac{p'}{p}\right)^{n+1} V_{i,\text{max}}(p,p')\text{ if } p>p',
	\label{Eq:Delta_p_V_i}
\end{align}
where the constants $\mathcal{M}_{V_i}$, $\mathcal{M}_{\Delta V_i,n}$ are dimensionless parameters of order one.
In the above inequalities, $V_i(\vec p\,',\vec p)$ are regarded as functions of the scalar quantities $p$ and $p'$.
The subtraction point is chosen to be $p=0$, which is not necessary, but makes the estimates simpler.

The important feature of the above inequalities is suppression of large momenta,
which is crucial for renormalizability of a theory.
This property survives the partial-wave projection, yielding the same bounds for the
partial-wave potentials as in Eq.~\eqref{Eq:Delta_p_V_i} with the simplification
that some terms in the expressions for the remainders vanish  
for higher partial waves due to the centrifugal barrier while others 
vanish due to parity conservation.

\section{Toy models}\label{Sec:ToyModels}
In this section we construct a general setting
based on simple separable interactions.
Various toy models that we analyze result as the limiting cases.

\subsection{Toy-model potentials}
Our toy models are motivated by chiral EFT interactions in the $NN$ sector.
We consider a system of two nucleons of mass $m_N$ in the $S$-wave.
The partial-wave potential is expanded in powers of the small scale $Q$:
\begin{align}
	V = V^{(0)} + V^{(2)} + ... \, .
	\label{Eq:PotentialExpansion}
\end{align}
The bare potential at each EFT order $i$ can be split into the renormalized part $V_i$ and the counter terms $\delta V_i$
\begin{align}
	V^{(i)} = V_i+ \sum_{j>i}\delta V_i^{(j)},
	\label{Eq:coutner_terms}
\end{align}
where the superscript $j$ means that the counter term $\delta V_i^{(j)}$ absorbs the power-counting breaking 
contributions, which formally appear at order $j$.
The LO ($\mathcal{O}(Q^0)$) and NLO ($\mathcal{O}(Q^2)$) potential consists of the short-range and long-range parts:
\begin{align}
	V_0(p',p)&=V_{0,\text{short}}(p',p)+V_{0,\text{long}}(p',p),\nonumber\\
	V_2(p',p)&=V_{2,\text{short}}(p',p)+V_{2,\text{long}}(p',p).
\end{align}
The short-range LO potential is a momentum-independent contact interaction,
regularized by a power-like form factor:
\begin{align}
	V_{0,\text{short}}(p',p) &=  V_{NN} C_0 F_{\Lambda}(p') F_{\Lambda}(p),\qquad F_{\Lambda}(p)=\frac{\Lambda^2}{\Lambda^2 + p^2},
\end{align}
where $V_{NN}=8\pi^2/(m_N\Lambda_b)$ is the overall normalization scale and $C_0$ is a dimensionless constant of order one.
The cutoff $\Lambda$ is assumed to be of the order of the hard scale $\Lambda\sim\Lambda_b$.
The long-range part mimics the  logarithmic branch point at
$p^2=-M_\pi^2/4$ of the partial-wave projected one-pion-exchange
potential with a pole located at $p^2=-M_\pi^2/4$,
\begin{align}
	V_{0,\text{long}}(p',p) &=  V_{NN} \Big\{ g_{1\pi} \big[F_{\Lambda}(p') F_{1\pi}(p)+F_{1\pi}(p') F_{\Lambda}(p)\big]+\tilde g_{1\pi} F_{1\pi}(p') F_{1\pi}(p)\Big\},
	\qquad F_{1\pi}(p)=\frac{M_\pi^2}{M_\pi^2 + 4\, p^2},
\end{align}
where the constants $g_{1\pi}$ and $\tilde g_{1\pi}$ determine
strength of the corresponding potentials.  
The two terms in $V_{0,\text{long}}$ correspond to the singular (appearing in spin-triplet $NN$ channels)
and regular (appearing in spin-singlet $NN$ channels) parts of the
one-pion-exchange potential. 
It is convenient to express the LO potential in the matrix form
\begin{align}
	V_0(p',p) &=  V_{NN} f_{0}^\intercal(p') v_0 f_{0}(p)= V_{NN} \sum_{i,j=1}^{2} f_{0,i}(p') v_{0,ij} f_{0,j}(p),
	\label{Eq:V_0_matrix_form}
\end{align}
with
\begin{align}
	v_{0} &= 
	\begin{pmatrix}
		C_0 & g_{1\pi}   \\
		g_{1\pi} & \tilde g_{1\pi}   
	\end{pmatrix},\qquad f_0(p)=(F_\Lambda(p), F_{1\pi}(p)).
	\label{Eq:v_0_matrix}
\end{align}

The potential $V_{2,\text{short}}$ is given by a contact interaction quadratic in momenta:
\begin{align}
	V_{2,\text{short}}(p',p) &=  V_{NN} C_2 \frac{p'^2 + p^2}{\Lambda_b^2} F_{\Lambda}(p') F_{\Lambda}(p).
\end{align}
The long-range part of the NLO potential features poles at $p^2 =
-M_\pi^2$, $p'{}^2 =
-M_\pi^2$ (motivated by the left-hand cut of the two-pion-exchange
potential that starts at the on-shell momentum $p^2=-M_\pi^2$):
\begin{align}
	V_{2,\text{long}}(p',p) &=  V_{NN} \, g_{2\pi}  \frac{p'^2 + p^2}{\Lambda_b^2}\big[F_{\Lambda}(p') F_{2\pi}(p)+F_{2\pi}(p') F_{\Lambda}(p)\big],
	\qquad F_{2\pi}(p)=\frac{M_\pi^2}{M_\pi^2 + p^2}.
\end{align}
The matrix form of the full NLO potential reads:
\begin{align}
	V_2(p',p) &= V_{NN} f_{2}^\intercal(p') v_2 f_{2}(p)=V_{NN} \sum_{i,j=1}^{4} f_{2,i}(p') v_{2,ij} f_{2,j}(p),
	\label{Eq:V_2_matrix_form}
\end{align}
with
\begin{align}
	v_{2} &=
	\begin{pmatrix}
		0& C_2 & 0 & g_{2\pi}   \\
		C_2& 0  & g_{2\pi} & 0   \\
		0& g_{2\pi}  & 0 & 0   \\
		g_{2\pi}& 0  & 0 & 0  
	\end{pmatrix},\qquad f_2(p)=\Big(F_\Lambda(p),\frac{p^2}{\Lambda_b^2}F_\Lambda(p), F_{2\pi}(p),\frac{p^2}{\Lambda_b^2}F_{2\pi}(p)\Big).
	\label{Eq:v_2_matrix}
\end{align}

\subsection{Solution of the Lippmann-Schwinger equation}
\subsubsection{Leading order}
To obtain the LO scattering amplitude, we solve the partial-wave
projected Lippmann-Schwinger equation, $T_0=V_0+V_0GT_0$, or, explicitly:
\begin{align}
	T_{0}(p',p;p_\text{on})&=V_0(p',p)+
	\int \frac{p''^2 dp''}{(2\pi)^3}
	V_0(p',p'')
	G(p'';p_\text{on})
	T_{0}(p'',p;p_\text{on}),\nonumber\\
	G(p''; p_\text{on})&=\frac{m_N}{p_\text{on}^2-p''^2+i \epsilon}.
	\label{Eq:LS_equation}
\end{align}
Its general solution is
\begin{align}
	&T_0=V_0 R=\bar R V_0,\label{Eq:T0_NP}
\end{align}
where $R$ ($\bar R$) is the resolvent of the Lippmann-Schwinger kernel
\begin{align}
	R=(\mathds{1}-G V_0)^{-1},\qquad \bar R=(\mathds{1}-V_0 G)^{-1}.
	\label{Eq:resolvent}
\end{align}
For the LO potential in the form~\eqref{Eq:V_0_matrix_form}, we have
\begin{align}
	T_0(p', p; p_{\text{on}}) =&V_{NN}  f_{0}^\intercal(p') t_{0}(p_{\text{on}}) f_{0}(p), \label{LOTmatrix}
\end{align}
where
\begin{align}
	t_{0}(p_{\text{on}}) =&v_{0} \, r(p_\text{on})= r^\intercal(p_\text{on}) v_{0},\qquad r(p_\text{on})= \left[\mathrm{I} - \Sigma_{0}(p_{\text{on}})v_{0}  \right]^{-1}.
\end{align}
The loop functions $\Sigma_0$ are defined as
\begin{align}
	\Sigma_{0,ij}(p_{\text{on}}) &= V_{NN}\int \frac{p''^2 \text{d}p''}{(2 \pi)^3}  f_{0,i}(p'') G(p'';p_\text{on}) f_{0,j}(p'') . \label{Eq:Sigma0}
\end{align}
Their explicit values are given by
\begin{align}
	\Sigma_{0,11}(p_{\text{on}}) &=\frac{\Lambda^3}{4\Lambda_b(p_\text{on}+i\Lambda)^2},\nonumber\\
	\Sigma_{0,12}(p_{\text{on}})&=\Sigma_{0,21}(p_{\text{on}})=-\frac{i \Lambda ^2 M_{\pi }^2}{2 \Lambda _b \left(2 \Lambda +M_{\pi }\right) \left(M_{\pi }-2 i p_{\text{on}}\right) \left(p_{\text{on}}+i \Lambda \right)},\nonumber\\
	\Sigma_{0,22}(p_{\text{on}}) &=-\frac{M_{\pi }^3}{8 \Lambda _b \left(M_{\pi }-2 i p_{\text{on}}\right){}^2}.
	\label{Eq:Sigma0_ij}
\end{align}
As one can see, $\Sigma_{0,11}$ is of order $\mathcal{O}(Q^0)$, whereas the other matrix elements of  
$\Sigma_0$ are of  order $\mathcal{O}(Q^1)$.

We can write down $t_0$ explicitly in the $N/D$ form
\begin{align}
	t_{0}(p_{\text{on}}) =& \frac{N(p_\text{on})}{D(p_\text{on})},
\end{align}
with
\begin{align}
	N(p_\text{on})=\left(
	\begin{array}{cc}
		C_0+g_{1\pi}^2 \Sigma_{0,22}-C_0 \tilde g_{1\pi} \Sigma_{0,22} & g_{1\pi} + C_0 \tilde g_{1\pi} \Sigma_{0,12}-g_{1\pi}^2 \Sigma_{0,12} \\
		g_{1\pi} + C_0 \tilde g_{1\pi} \Sigma_{0,21}-g_{1\pi}^2 \Sigma_{0,21} & \tilde g_{1\pi}+g_{1\pi}^2 \Sigma_{0,11}-C_0 \tilde g_{1\pi} \Sigma_{0,11} \\
	\end{array}
	\right),
\end{align}
and the Fredholm determinant 
\begin{align}
D(p_\text{on})=\det[\mathds{1}-\Sigma_{0}v_0].
\end{align}
Both $N$ and $D$ are quantities of order one, i.e., $\mathcal{O}(Q^0)$.
Therefore, $t_0$ (and $T_0$) is of order $\mathcal{O}(Q^0)$, as expected, unless
$D(p_\text{on})$ is unnaturally small, which leads to an enhancement of the LO amplitude
similarly to the situation in the $NN$ $S$-wave channels.
The matrix $r$ can also be represented as the ratio
\begin{align}
	r(p_\text{on})=\frac{Y(p_\text{on})}{D(p_\text{on})},	
\end{align}
with
\begin{align}
	Y
	&=\left(\begin{array}{cc}
		1-g_{1\pi}  \Sigma_{0,21}-\tilde g_{1\pi}  \Sigma_{0,22} & g_{1\pi}  \Sigma_{0,11}+\tilde g_{1\pi}  \Sigma_{0,12}   \\
		C_0  \Sigma_{0,21}+g_{1\pi}  \Sigma_{0,22} & 1-C_0  \Sigma_{0,11}-g_{1\pi}  \Sigma_{0,12}
	\end{array}\right),
\end{align}
and $Y(p_\text{on})=\mathcal{O}(Q^0)$.
\subsubsection{Next-to-leading order}
The NLO amplitude is calculated perturbatively to explicitly control the contributions of different EFT orders.
The unrenormalized NLO amplitude is given by
\begin{align}
	&T_2=\bar R V_2 R\label{Eq:T2_NP}.
\end{align}
For the separable potential defined in Eq.~\eqref{Eq:V_2_matrix_form}, we obtain
\begin{align}
	T_2(p',p;p_{\text{on}}) 
	=& V_{NN} \psi_2^\intercal(p_{\text{on}}) v_2 \psi_2(p_{\text{on}}),
	\label{NLOamplitude}
\end{align}
where $\psi_2$ is given by
\begin{align}
	\psi_2(p_{\text{on}}) =&\,f_2(p_{\text{on}}) + \Sigma_2(p_{\text{on}}) t_0(p_{\text{on}}) f_0(p_{\text{on}}),
\end{align}
with
\begin{align}
	\Sigma_{2,ij}(p_{\text{on}})&= V_{NN}\int \frac{p''^2 \text{d}p''}{(2 \pi)^3} f_{2,i}(p'')G(p'';p_\text{on})
	f_{0,j}(p''). \label{Eq:Sigma2}
\end{align}
The explicit form of the matrix elements of $\Sigma_2$ are given by
\begin{align}
	\Sigma_{2,11}(p_{\text{on}}) &=\Sigma_{0,11}(p_{\text{on}}),\qquad 
	\Sigma_{2,12}(p_{\text{on}}) = \Sigma_{0,12}(p_{\text{on}}),\nonumber\\
	\Sigma_{2,21}(p_{\text{on}}) &=\frac{\Lambda ^4 \left(\Lambda -2 i p_{\text{on}}\right)}{4 \Lambda _b^3 \left(p_{\text{on}}+i \Lambda \right)^2},\nonumber\\
	\Sigma_{2,22}(p_{\text{on}}) &=-\frac{\Lambda ^2 M_{\pi }^2 \left(2 \Lambda  p_{\text{on}}+M_{\pi } \left(p_{\text{on}}+i \Lambda \right)\right)}{4 \Lambda
		_b^3 \left(2 \Lambda +M_{\pi }\right) \left(M_{\pi }-2 i p_{\text{on}}\right) \left(p_{\text{on}}+i \Lambda \right)},\nonumber\\
	\Sigma_{2,31}(p_{\text{on}}) &=-\frac{i \Lambda ^2 M_{\pi }^2}{2 \Lambda _b \left(\Lambda +M_{\pi }\right) \left(M_{\pi }-i p_{\text{on}}\right) \left(p_{\text{on}}+i \Lambda \right)},\nonumber\\
	\Sigma_{2,32}(p_{\text{on}}) &=-\frac{M_{\pi }^3}{6 \Lambda _b \left(-3 i M_{\pi } p_{\text{on}}+M_{\pi }^2-2 p_{\text{on}}^2\right)},\nonumber\\
	\Sigma_{2,41}(p_{\text{on}}) &=-\frac{\Lambda ^2 M_{\pi }^2 \left(\Lambda  p_{\text{on}}+M_{\pi } \left(p_{\text{on}}+i \Lambda \right)\right)}{2 \Lambda _b^3 \left(\Lambda +M_{\pi }\right) \left(M_{\pi
		}-i p_{\text{on}}\right) \left(p_{\text{on}}+i \Lambda \right)},\nonumber\\
	\Sigma_{2,42}(p_{\text{on}}) &=-\frac{M_{\pi }^4 \left(M_{\pi }-3 i p_{\text{on}}\right)}{12 \Lambda _b^3 \left(M_{\pi }-i p_{\text{on}}\right) \left(M_{\pi }-2 i p_{\text{on}}\right)}.
	\label{Eq:Sigma2_ij}
\end{align}
One expects from the power counting that the NLO amplitude is of order $\mathcal{O}(Q^2)$.
However, there are terms in $T_2$ that violate the power counting.
They are proportional to $\Sigma_{2,21}$ and originate from the
integrals containing in the integrand the factor $p^2/\Lambda_b^2$, which is 
naively of the order of $Q^2$, but leads to 
the factor $\Lambda^2/\Lambda_b^2$, i.e., $\mathcal{O}(Q^0)$, 
if the integral converges at momenta $p\sim\Lambda$,
Such PCB terms are linear in $\Sigma_{2,21}$ due to the off-diagonal structure of $v_2$.
To remove the PCB contribution, we introduce the counter term $\delta V_0^{(2)}$,
so that the renormalized expression for the NLO amplitude becomes
\begin{align}
	&\mathds{R}(T_2)=\bar R \left(V_2 +\delta V_0^{(2)}  \right)R=T_2+\delta T_2,\qquad \delta T_2=\bar R \delta V_0^{(2)}R.\label{Eq:T2_renormalized}
\end{align}

The on-shell counter-term contribution reads
\begin{align}
	\delta T_2(p_{\text{on}})&= V_{NN} \psi_0^\intercal(p_{\text{on}}) \delta v_0\, \psi_0(p_{\text{on}}),
\end{align}
with
\begin{align}
	\psi_0(p_{\text{on}})=r(p_{\text{on}}) f_0(p_{\text{on}}).
\end{align}
Note that
\begin{align}
	\psi_{2,1}(p_{\text{on}})=\psi_{0,1}(p_{\text{on}}),
\end{align}
because $f_{2,1}(p)=f_{0,1}(p)$.
In the next section, we consider various particular cases when the power-counting breaking terms
can or cannot be absorbed by the counter-terms.

\section{Renormalization of the NLO amplitude}\label{Sec:Renormalization}
We start by listing the power-counting breaking contributions in $T_2$ proportional to $\Sigma_{2,21}$:
\begin{align}
	T_2^\text{PCB,I}(p_{\text{on}}) 
	&=V_{NN}\psi_{0}^\intercal(p_{\text{on}}) X^\text{PCB,I}(p_{\text{on}}) \psi_{0}(p_{\text{on}}),\nonumber\\
	X^\text{PCB,I}_{ij}(p_{\text{on}})&=\delta_{i1} v_{2,12}  \Sigma_{2,21}(p_{\text{on}}) v_{0,1j}
	+  v_{0,i1} \Sigma_{2,21}(p_{\text{on}})v_{2,21} \delta_{j1},\label{Eq:T2_PCB_I}
\end{align}
and
\begin{align}
	T_2^\text{PCB,II}(p_{\text{on}}) 
	&=V_{NN}\Big[\psi_{2}^\intercal(p_{\text{on}}) X^\text{PCB,II}(p_{\text{on}}) \psi_{0}(p_{\text{on}})
	+\psi_{0}^\intercal(p_{\text{on}}) \big(X^\text{PCB,II}(p_{\text{on}})\big)^\intercal \psi_{2}(p_{\text{on}})\Big],\nonumber\\
	X^\text{PCB,II}_{ij}(p_{\text{on}})&=\delta_{i3} v_{2,32}  \Sigma_{2,21}(p_{\text{on}}) v_{0,1j}.\label{Eq:T2_PCB_II}
\end{align}

Now, we consider four different models by switching on and off various terms in the LO and NLO potentials.

The first model consists of purely short-range interactions with  $g_{1\pi}=\tilde g_{1\pi}=g_{2\pi}=0$.
Then, the only PCB term 
comes from
\begin{align}
	X^\text{PCB,I}_{11}(p_{\text{on}})&= 2v_{2,12}  \Sigma_{2,21}(p_{\text{on}}) v_{0,11}=
	2\,C_2 C_0 \Sigma_{2,21}(p_{\text{on}}).
\end{align}
The PCB term can be removed from $\Sigma_{2,21}$ by just subtracting
its value at threshold $\Sigma_{2,21}(0)=-\Lambda^3/(4\Lambda_b^3) $:
\begin{align}
	\delta v_{ij}= C_2C_0\frac{\Lambda^3}{2\Lambda_b^3} \delta_{1i}\delta_{1j}.
\end{align}
The remaining (renormalized) part of the amplitude is of order $\mathcal{O}(Q^2)$ since
\begin{align}
\Delta \Sigma_{2,21}(p_{\text{on}})=	\Sigma_{2,21}(p_{\text{on}})-\Sigma_{2,21}(0)=-\frac{\Lambda ^3 }{4 \Lambda _b \left(\Lambda -i p_{\text{on}}\right)^2}\frac{p_{\text{on}}^2}{\Lambda _b^2}
	=\mathcal{O}(Q^2).
\end{align}
This result is, of course, well known. In this case, the interaction
trivially fulfills the 
locality condition of the long-range interaction as there is no long-range interaction at all.
Interestingly, the same counter term renormalizes the theory with $\tilde g_{1\pi}\ne 0$ ($g_{1\pi}=g_{2\pi}= 0$)
because, in this case, there is no off-diagonal terms in the LO potential.
The only quantity that changes is $\psi_0(p_{\text{on}})$.
This is analogous to the situation in the spin-singlet channels in the $NN$-system,
where the one-pion-exchange potential is regular.
Thus, in this case, even though the long-range potential is nonlocal, the theory 
is still renormalizable because no new PCB terms arise.

Consider now the model with the singular LO "one-pion-exchange" potential, i.e.
$\tilde g_{1\pi}=g_{2\pi}=0$, $g_{1\pi}\ne 0$.
Then, there are additional PCB contributions due to
\begin{align}
	X^\text{PCB,I}_{12}(p_{\text{on}})&=X^\text{PCB,I}_{21}(p_{\text{on}})=v_{2,12}  \Sigma_{2,21}(p_{\text{on}}) v_{0,12}=
	2\,C_2 \,g_{1\pi}\, \Sigma_{2,21}(p_{\text{on}}).
\end{align}
To absorb these terms, we need to introduce the counter terms with constants
\begin{align}
	\delta v_{12}=\delta v_{21}=C_2 \,g_{1\pi}\frac{\Lambda^3}{4\Lambda_b^3}.
\end{align}
The full counter-term potential, therefore, has the following
form:
\begin{align}
	\delta V(p',p)= C_2 \frac{\Lambda^3}{4\Lambda_b^3}\big\{2C_0 f_{0,1}(p')f_{0,1}(p)+g_{1\pi}\big[f_{0,1}(p')f_{0,2}(p)+f_{0,2}(p')f_{0,1}(p)\big]\big\},
\end{align}
and contains the long-range part.
This is in contrast with chiral EFT, where the counter terms are local and, therefore, short-range.
This fact was also noticed in Refs.~\cite{Epelbaum:2017byx,Peng:2021pvo}.
We conclude that this model can be renormalized by means of introducing a nonlocal counter term,
nevertheless, without modifying the structure of the LO potential.

Finally, we consider the case with the non-vanishing long-range NLO potential
corresponding to the two-pion exchange, i.e. $g_{2\pi}\ne 0$.
The term $T_2^\text{PCB,II}$ in Eq.~\eqref{Eq:T2_PCB_II} contains the component
$\psi_{2,3}(p_{\text{on}})$, which does not reduce to $\psi_0$ as in all above cases.
The PCB term has again a long-range structure ("two-pion-exchange" singularity).
However, this kind of structure is not present in the LO potential, so one cannot
absorb it by redefining any LO LEC.
Therefore, unless we introduce a "two-pion-exchange" term of order $\mathcal{O}(Q^0)$ in the LO potential, 
the renormalization cannot be performed. 
Obviously, including all possible structures (required to absorb PCB terms) 
into the LO potential contradicts the idea
of a systematic expansion of the amplitude in the spirit of an EFT.

To summarize, we have demonstrated that if the condition of locality of the long-range forces
is violated, renormalization might either require the introduction of non-local counter terms
or even become impossible.

\section{Summary}\label{Sec:summary}
We have analyzed several toy models featuring separable interactions
motivated by the nucleon-nucleon chiral EFT potential.
We considered the leading-order ($\mathcal{O}(Q^0)$)
and next-to-leading-order ($\mathcal{O}(Q^2)$) potentials in the
EFT expansion.
Those interactions contain the short-range parts, i.e. the contact terms,
and the long-range parts corresponding to the one- and two-pion-exchange contributions.
The LO interaction was treated nonperturbatively by solving the Lippmann-Schwinger equation.
A nonlocal regulator with a cutoff of the order of the hard scale $\Lambda\sim\Lambda_b$
was introduced to render all integrals finite.
The NLO amplitude was then calculated perturbatively to analyze individual contributions
to various EFT orders.

The goal of our study was to check whether those models are renormalizable.
That means that the power-counting breaking contributions in the NLO amplitude proportional to positive powers of
$\Lambda$ and stemming from the integration momenta $p\sim\Lambda$ can
be absorbed via a redefinition of the LO potential.
The reason for questioning the renormalizability feature is 
the fact that
the models considered do not satisfy one of the sufficient
renormalizability criteria, stating that the long-range part of the interaction
must be local. 

The first model we analyzed does not contain any long-range interaction.
Therefore, as expected, it turns out to be renormalizable.
Adding to this model a regular (decreasing at large momenta) long-range LO
interaction does not affect its renormalizability since no new power-counting breaking
integrals appear.

If we instead add a singular long-range LO potential motivated by the 
one-pion-exchange term in the spin-triplet $NN$ channels,
the model cannot be renormalized in a standard way by introducing
a local counter term in the LO potential.
Nevertheless, renormalization is still possible with a long-range
counter term without modifying the structure of the LO potential.

Finally, we analyzed the model with the separable long-range NLO potential
motivated by the two-pion-exchange contribution in chiral EFT.
In this case, renormalization cannot be realized unless one includes
the same kind of structure in the LO potential.
This means that the LO potential must contain all possible
types of interactions appearing at higher order, which is
obviously at odds with the idea of an effective field theory.

Thus, we have demonstrated that the renormalizability of effective
theories of nuclear interactions cannot be taken for granted when using
arbitrary phenomenological models, but represents a specific feature of
interactions based on an EFT.

\section*{Acknowledgments} 
We thank Jambul Gegelia for helpful comments on the manuscript.
This work was supported by DFG (Grant No. 426661267),
by the European Research Council (ERC) under the European Union's
Horizon 2020 research and innovation programme (grant agreement
No. 885150) and by the MKW NRW under the funding code NW21-024-A.

\bibliography{6.0}
\bibliographystyle{apsrev}

\end{document}